Do Quantum Systems Break The Equivalence Principle?

Timir Datta

Physics and Astronomy department

University of South Carolina

Columbia, SC 29208

And

Ming Yin

Physics/Engineering Department

Benedict College

Columbia, SC 29204

**Overview:**

Gravitational response of real objects is a fascinating topic. Einstein formalized the Galileo-Newton ideas of equality of free falls into "the complete physical equivalence" or the Principle of Equivalence [Albert Einstein, *The meaning of Relativity*, 5$^{th}$ ed. Princeton, (1921)]. This principle (EP) introduced physical content into the mathematical postulate of general covariance [W. Pauli, *Theory of relativity*, Dover, New York (1981)] and led to General Relativity.

However, in this article we point out that in a gravitational field, *g*, the bulk response of an electrically neutral but atomistic test mass is model dependant. Depending on the particular quantum approximation scheme, opposing results for the gravity induced (electric) polarization $P_g$ have been reported. For instance, $P_g$ is small and oriented anti-parallel to *g*, if the deformations of the positive background lattice is neglected [L.I. Schiff, PRB, **1**, 4649 (1970)]. But, it is $10^5$ larger and opposite in direction in the elastic lattice approximation [A. J. Dessler et al, Phys.Rev, **168**, 737, (1968); Edward Teller, PNAS, **74**, 2664 (1977)]. Hence, the elastic model contradicts reports of polarization in accelerated metals [Richard C. Tolman & T.Dale Stewart, Phys Rev **28**, 794 (1926); G. F. Moorhead & G. I. Opat, Class. Quant. Grav, **13**, 3129 (1996)]. Surprisingly, the rigid system is consistent with EP but the elastic system breaks EP. Here the historical literature is surveyed and some implications are outlined.

There has been a human fascination with movement, motion, the lack of motion (equilibrium) and weight (gravitation), through out history[1-2]. As evidenced by the great pyramids, already by the time of the Pharaonic dynasties an immense amount of mechanical knowledge was at hand. The "Egyptian's" mastery of out-of- balanced forces is clear from the design of the Anubis' balance[3] (Figure 1). The extremely long suspensions of the pans compared with the beam length indicate the instrument's great sensitivity.

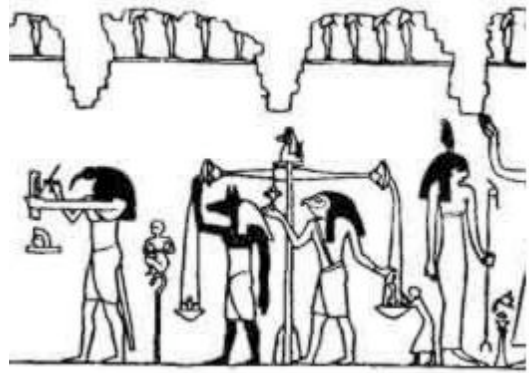

**Figure 1:** Egyptian scene depicting a highly sensitive balance.

The sophistication of classical Greco-Roman civil engineering indicates continued growth of knowledge and technology. However, recorded history of this period has been dominated by the polemics of "Aristotle et al" and is much too well known to repeat here. Instead we will mention some of the less publicized contributions.

Unfortunately, most of the pre and proto-science is lost to history. The surviving written records begin roughly at the start of the Common Era (CE). Amongst these the work of Aryabhata (476–550 CE) the Indian mathematician and astronomer, indicate clear breaks from the Aristotle - Ptolemy (AP) world view. About the same period, Egyptians scholars regained eminence mostly Greek speaking, particularly *John*

*Philoponus* (aka John the Grammarian 490-570 CE) an Alexandrian like Ptolemy and a leading Aristotelian commentator also refuted some of the AP ideas.

But it was in Europe during renaissance, especially in the sixteenth century that natural history was thoroughly updated. Regardless of the verity of the folklore about Galileo's (1590) *experimentum crucis* of simultaneously dropping a cannon ball and a musket shot from the top of the leaning tower in Pisa and showing "Aristotle wrong".

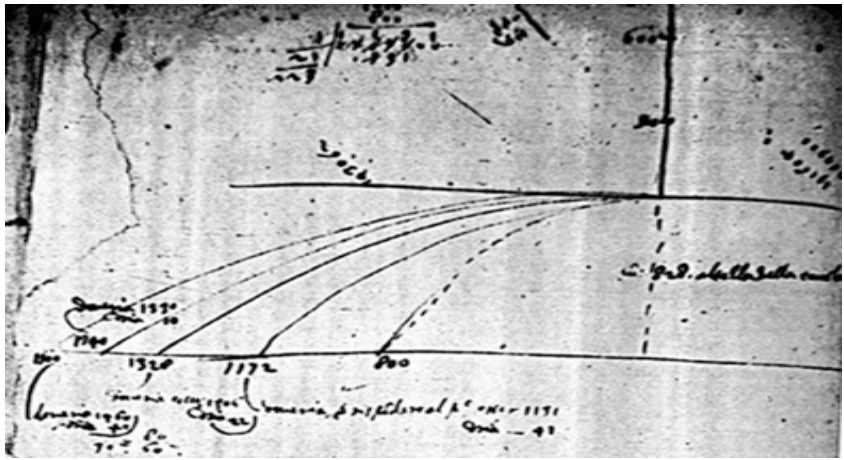

Figure 2: A page from Galileo's diary depicting trajectories of a projectile with different initial velocities.

Historically, the accolade for equality of time of free flight belongs if not to the (1553) writings of Benedetti Giambattista then surely to Simon Stevin (1586) the Flemish engineer and discoverer of the hydrostatic paradox. But from Galileo's note book drawings [4], there can be no doubt about his correct understanding of projectile motions. In particular as to what are the determining parameters and how marginal is the role of mass (Figure 2).

However, most credit is due to Isaac Newton. Who, perhaps after Robert Hook's suggestion as to the direction and inverse proportionality of gravitational force, overcame his abhorrence to action at a distance (concept of field) and fully appreciated Galileo's amanuensis and assistant Vincenzio Viviani's conclusion regarding "the equality of fall rates" [5-8].

Descent of real objects towards the earth can be far more complicated. To account for peculiarities in motion inside dense liquids, material specific forces due to Archimedes, Stoke and others were required. Historically, it was only after the invention of the air (vacuum) pump by Robert Boyel and Edme Marriotte's famous "guinea-and-feather" experiment that the role of the medium (atmospheric air) can be demonstrated[1,2].

Albert Einstein[9] extended the equality of the rate of fall by introducing the "…the assumption …of the principle of equivalence" (EP). EP transformed the mathematical postulate of general covariance into a physical theory of gravity[10]. Currently several versions of EP are in circulation but here we follow Einstein's remarkably original, limpid and succinct enunciation[9]. It really is remarkable that - a "flower of Kent" hanging in a Lincolnshire apple orchard, circulating beam of relativistic elementary particles at CERN[11], the detectors at the Laser Interference Gravity Observatory (LIGO) [13], or the laser cooled cesium atoms in Steven Chu's laboratory[11] and in the Canadian, NRC fountain clock[14] all respond to gravity the same way. Credits for championing the equality between gravitational and inertial mass also belong to Huygen, Bessel, Eotvos and more recently to Dickie[15,16].

In the ultimate analysis all things are atomic and quantum. It is fair to ask how a quantum object responds to gravity. Remarkably, this question of gravitational response of a bulk quantum system is not as clear. Even in the non-relativistic limit, due to the diverse requirements of Fermi-statistics, exchange and correlation energies the problem remains complex [17]. For instance, as shown by Wigner[18] even the ground state of the electron gas is non-trivial and of broken symmetry[19]. Equilibrium of a solid crystal and conduction electrons under gravity is particularly challenging; the problem requires delicate quantum approximations. Here we survey the important literature and propose a direct experiment to sort out some of the questions regarding gravitational response of a small but macroscopic test mass made of real matter.

The first observations on inertia of electric charges were in late 1800's pertaining to electrolytes. Tolman[20] and his colleagues at Caltech were the pioneers in inertial behaviors of electrical charges in conducting solids. They argued that the rear (trailing) end of a linearly accelerating metal rod will be negatively charged "owing to the lagging behind of the relatively mobile electrons" and like wise the circumference of a rotating disk will also be negatively charged. They reported observing these (small) effects. Acceleration experiments were repeated by Beams[21] at the University of Virginia and elsewhere by others[22-28]. In summery, as shown in figure 3, under either linear or radial acceleration the conduction electrons appear to behave as expected-similar to that of a classical fluid spinning in Newton-Mach's water pail.

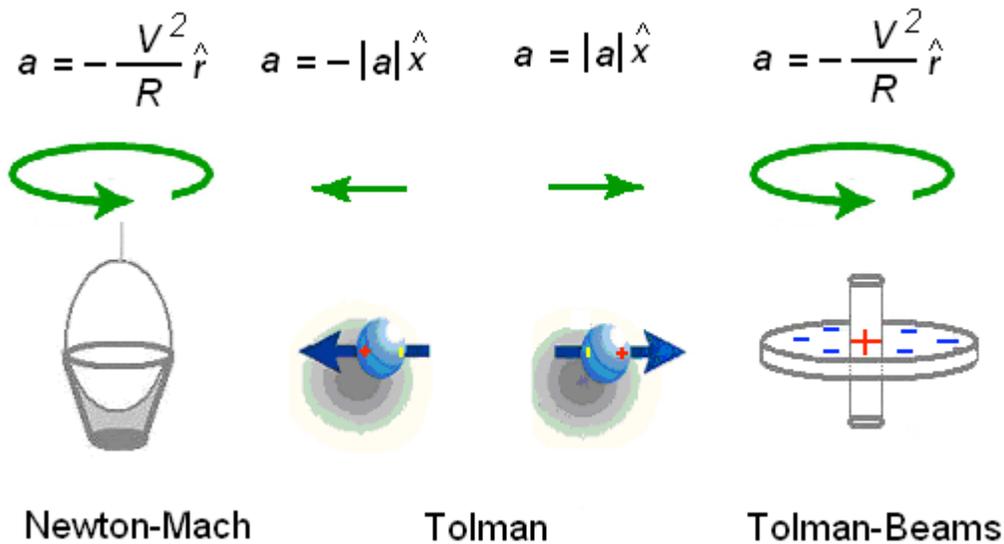

**Figure 3:** Effects of linear and radial acceleration on classical and quantum fluids inside a conductor.

Understanding charges and atomic matter takes us well into the twentieth century. During the development of theory of electromagnetism, the nature of electric charge and its response to forces were much debated. These led to the discovery of Hall Effect by Edwin Hall (1879) leading to the conclusion that contrary to Maxwell the charge carriers (not the metal) are the recipient of the (Lorentz) force. About twenty years later (1897) J.J. Thomson identified these charges to be the electrons, particles carrying fundamental unit of electric charge proposed by George Johnstone Stoney three years earlier. Thompson also promulgated the "plum-pudding" atom favored amongst others by Lord Kelvin. In 1900 Paul Drude produced the eponymous free particle model for charge conduction. And in 1909 after the Geiger-Marsden experiment Rutherford introduced his model where the atomic volume is mostly filled with orbiting negative charges and the positive charges are densely concentrated at the massive but tiny regions, the nucleus. Rutherford's scheme rapidly dispatched the Thompson model, because the miniature solar system description perfectly suited the development of quantum mechanics by Bohr, Sommerfeld, Schrödinger and others.

With the advent of wave mechanics and advances in the quantum many body description a better modeling of the negatively charged Landau quasi particle (electrons) of the Fermi-fluid and the positive atomic cores in metals was possible. When the Fairbank[29] group at Stanford started precision experiments of electron free fall, there was an incentive to provide theoretical description of conductors under gravity[30-32]. Schiff also at Stanford proposed a model Hamiltonian for a conductor in the gravity field $g$, is considered to be:

$$H = H_0 + H_g + V \qquad \ldots 1$$

where, $H_o$ is the system without gravity and $V$ represents the supporting constraints. $H_g$ is the gravitational potential energy measured from the x-y plane, i.e.

$$H_g = g[m_e \sum_i z_i + M_c \sum_j z_j] \qquad \ldots 2$$

Here, g is the free fall acceleration, $m_e$ the electron mass and $M_c$ is the core mass. Calculating the equilibrium field of the above Hamiltonian requires a number of approximations. Schiff essentially argued that the degrees of freedom associated with massive cores can be neglected and only the conduction electrons redistribute under the gravitational force. This force will cause the electrons to sink towards the bottom so that an internal electric field, $E_g$, pointing vertically down, builds up. Notice, the seeds of the "future' are already inherent in this reduced Hamiltonian of equation 1; because in this break down the effect of gravity is just the additional constraint of the "support" and the external potential energy term $H_g$.

Shortly after the publication of the first paper, a controversy started over the incorrect assessment of the lattice contribution in the rigid model. Calculations by Dessler etal[33] and others included better accounting of lattice compressibility effect. Physically, on an elastic system earth's gravity produces a far greater compression of the massive lattice creating a bigger positive charge density (background) compared with that of the far lighter and less compressible (Fermionic) free-electrons. The conductor as whole gets polarized with the positive pole at the bottom. Consequently the conductor would also acquire an electric dipole moment $\boldsymbol{P}_g$, which points vertically down.

These calculations predicted that the induced electric field is in the opposite direction of gravity that is up ward, and is much bigger. Indeed, with the inclusion of the core contributions $\boldsymbol{Eg} \sim g/q(M_c/m_e)$ where, g is the free fall acceleration, q is the unit of electronic charge. Herring took a thermodynamic approach[34]. In elastic models[33-36] the estimate for $\boldsymbol{E}_g$ come out to be the same order of magnitude ($\sim 10^5$) higher.

Teller considered[37] insulating dielectric matter with two different ionic masses, $M_+$ and $M_-$ respectively. From the calculations of electric dipole moment of a rapidly rotating dielectric Teller predicted the generation of magnetic fields near the object. Even for systems with the largest ionic mass ratio ($M_+/M_-$) it is not possible to reach the high value of ($M_c/m_e$) so very rapid rotation will be required to create large enough acceleration to produce detectable signals. He argued that such acceleration measurements can be important in the investigations of ferroelectric polarization and related phenomena. Unfortunately, surface field measurements of rotating objects are difficult especially at high angular velocities and the technique has received little attention but concerns about rotation on conductors remain active[21-28].

As described above, the difference between the perfectly rigid solid and the more realistic elastic models is manifestly quantitative- the effect is entirely due to the massive core lattice and five orders of magnitude larger in the elastic limit. However, the most striking difference between the two models is qualitative $\boldsymbol{E_g}$ are in opposite directions.

As a consequence of $E_g$ the whole object will acquire a gravitationally induced electric dipole moment $P_g$. **NB**, most researchers consider $E_g$ not $P_g$. This is important because unlike the intensive quantity $E_g$, the induced moment is an extensive quantity so a large volume objects give rise to proportionately larger signals. Even with elastic enhancement, in engineering terms the induced field is, it is rather small $E_g \sim$ 1 microV/m and hence it is better to measure $P_g$ rather than $E_g$. $P_g$ provides the advantage of cumulative built up of the effect of gravity over the total sample.

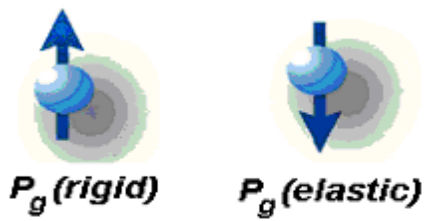

**Figure 4:** The direction of the gravitationally induced dipole moments in the rigid and elastic models.

In these two types of models $P_g$ is also predicted to be in opposite directions. It is parallel to gravity in the elastic case and in the anti-parallel direction in the rigid model as shown in figure 4. In the elastic case is correct then by locally observing the induced polarity of $P_g$ at one point on the surface of a (finite size $L<c^2/g$) conducting test mass it will be possible to distinguish between physical acceleration and static uniform gravity.

A pictorial comparison between the electric polarization of a finite size solid sphere under constant linear (vertical) acceleration and when placed in a vertically downward gravitational field are depicted in figure 5. The blue arrows point the direction of the induced electric dipole moments in each case. The left panel of shows the response under acceleration, while the center and right panels show the rigid and elastic behaviors under gravity Not surprisingly, rigid type behavior is consistent with the principle of equivalence. On the other hand the elastic lattice breaks Eisntein's equivalence principle[9].

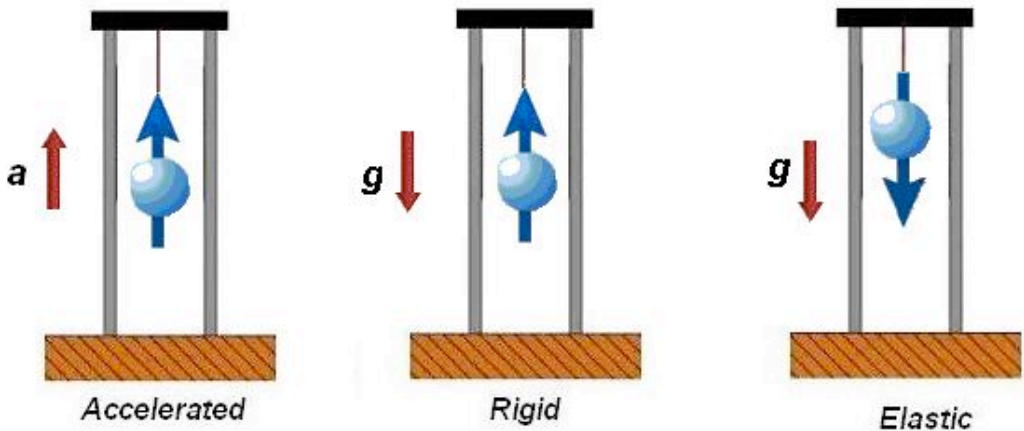

**Figure 5**: A comparison between the polarization of solid test mass in linear acceleration and under gravity.

An experiment to resolve the Schiff-Dessler, or Rigid and Elastic conductor controversy regarding strength and direction of $P_g$ has been proposed [38]. Such an experiment can impact several other topics, such as the Schiff gravity gyroscope and results of electric charge quantization measurements [39-41]. Also devices for acceleration-gravitation discriminating detectors and sensors for mapping or imaging the density distributions inside radiation sensitive soft objects such as the human body and others [42] are potential technological applications.

In the search for an answer to the basic question – does the effect of acceleration on real (quantum) matter "completely similar" to that in gravity? Also, is the total mass of a small object the only parameter in determining the full response under gravity? We have come a long way from Aristotle to electrons in a quantum solid. It was pointed out that even in a very weak uniform gravitational field, g, the behavior of a real object including those of dimensions far smaller than $c^2/g$ can be different from that of a homogeneous classical (point) particle. The disparate needs of the different quantum constituents inside the object have to be considered. Einstein's Principle of Equivalence is broken, i.e., the physical behavior of an object under constant acceleration will not be "completely similar" to that under uniform gravity.

In the quantum field theory even "perfect" empty (free) space has self-energy and produces dielectric screening. Hence, the free fall acceleration of test particles need not be independent of the particles internal structure. As a consequence it may be possible to distinguish accelerated motion from gravitation by a local observation.

**Conclusion:**

Here we have surveyed the literature for historical accuracy some of the less known researchers are noted. Bulk matter is atomic, atoms are composite and the gravitational response depends on the makeup of the atoms. Not surprisingly, the gravitational influence on matter made of "Thomson's atoms" need not be identical to that of Rutherford. Conducting matter composed of Rutherford's atoms follow quantum mechanics. As a consequence an external gravitational field *g*, even though far weaker than electromagnetism, can produce a net measurable response. In some ways this behavior is reminiscent of Hall Effect mentioned earlier, where in the net Lorentz-force, the magnetic contribution is only ~ 1/c of the electric force, but still it is the one that determines the Hall voltage.

A combination of screening, thermodynamics and conservation laws determine the bulk properties of a quantum system. But the details of the approximations are critical because different schemes can give rise to qualitatively different behaviors. As discussed above, if the elastic approximation is more realistic then even in a very weak uniform gravitational field, the complete physical response of a real conductor including those of dimensions far smaller than $c^2/g$ cannot be described as that of a homogeneous classical (point) particle. And are led to conclude that mass is not the only parameter of a system that determines gravitational behavior. Thus, it will be possible to distinguish accelerated motion from gravitation by a local observation.

**Acknowledgments:**

One of us (TD) would like to dedicate this work to the memories of William Fairbank and Jeeva Anandan; to Bill for encouraging experimental gravity during his many visits in the 1980's. To Jeeva for the innumerable debates we would have about gravitation in composite systems, the especially meaning of effective mass and inertia in solids. Thanks to Yakir Aharonov for many conversations regarding acceleration in quantum systems. Thanks to Alexei Abrikosov for remarks on the role of weak interactions and the effect of gravity on conducting systems. Also, very special acknowledgment to P. W. Anderson for his comments about electrons in metals, the perspective from the Bell Laboratories and references to the work that was followed Fairbank and Schiff.